\documentclass[preprint,prc,showpacs,preprintnumbers,amsmath,amssymb,floatfix]{revtex4-2}
\usepackage{amsmath,amsfonts,amssymb}
\usepackage{mathtools}
\usepackage{bm}

\usepackage{CJKutf8}
\usepackage{setspace}
\usepackage{graphicx}
\usepackage{booktabs}
\usepackage{float}
\usepackage{longtable}
\usepackage{multirow}
\usepackage{array}
\usepackage{makecell}

\newcolumntype{C}[1]{>{\centering\arraybackslash}p{#1}}

\usepackage[shortlabels]{enumitem}
\usepackage{xcolor}

\usepackage{listings}
\usepackage[colorlinks,linkcolor=blue,citecolor=blue,urlcolor=blue]{hyperref}

\begin{document}
			\title{Relativistic mean-field study of the neutron star inner crust using the asymmetric finite difference method}
			 \author{Jinzhe Zhang}
			\affiliation{School of Physics, Nankai University, Tianjin 300071,  China}

			 \author{Ying Zhang}
			 \thanks{Contact author:yzhangjcnp@tju.edu.cn}%
			\affiliation{Department of Physics, School of Science, Tianjin University, Tianjin 300072, China}
			
			\author{Jinniu Hu}
			\thanks{Contact author:hujinniu@nankai.edu.cn}%
			\affiliation{School of Physics, Nankai University, Tianjin 300071,  China}
			\affiliation{Shenzhen Research Institute of Nankai University, Shenzhen 518083, China}
			
			\author{Hong Shen}
			\affiliation{School of Physics, Nankai University, Tianjin 300071,  China}

			\begin{abstract}
				The ground-state properties of neutron-rich nuclear clusters in the inner crust of neutron stars are investigated within the Wigner–Seitz approximation using a relativistic mean-field framework. The radial Dirac equations are solved with an asymmetric finite-difference scheme, by which the hermiticity is preserved and spurious states are eliminated. Calculations are performed for representative Wigner–Seitz cells employing TM1-based interactions with different symmetry-energy slope parameters $L$, as well as a parametrization with a larger nucleon effective mass. It is found that the binding energy per nucleon decreases systematically with increasing $L$, while a larger effective mass leads to further reduction, particularly at higher densities. Quantum shell effects, which are absent in the Thomas–Fermi approximation, give rise to oscillatory density distributions and modify neutron properties. 
				Within the Wigner-Seitz cell, the resulting neutron root-mean-square radius and chemical potential
				are shown to be sensitive to both $L$ and the effective nucleon mass, underscoring their important roles in determining the microscopic structure of the neutron-star inner crust.				
			\end{abstract}

			\vspace{1.5cm}
			
			\maketitle
			
			\section{INTRODUCTION}
			Neutron stars represent the densest observable matter in the Universe and serve as unique natural laboratories for testing nuclear physics under extreme conditions~\cite{oertel2017equations,Haensel2007}. A typical cold neutron star is structured into three distinct regions beneath its atmosphere, the inhomogeneous outer crust, the inner crust, and a dense, homogeneous liquid core. The outer crust consists of a body-centered cubic (BCC) lattice of nuclei embedded in a degenerate electron gas, which ensures global charge neutrality. As the depth increases, the baryon number density increases accordingly, and the nuclei become progressively more neutron rich~\cite{lattimer2004physics,chamel2008physics,chamel2015brussels}. When the density reaches the neutron drip threshold, nuclei are no longer able to bind all neutrons~\cite{pearson2011properties,kreim2013nuclear}. This leads to the so-called neutron-drip phenomenon, which marks the transition to the inner crust. In this region, the nuclear matter is composed of extremely neutron-rich nuclear clusters immersed in a dilute gas of free neutrons and electrons. As the density further increases toward the crust–core transition, the interplay between nuclear surface tension and Coulomb repulsion favors the formation of complex non-spherical structures, commonly referred to as nuclear pasta phases~\cite{hashimoto1984shape}.
			
			{The matter in the inner crust of a neutron star can be regarded as a nearly perfect crystal and should, in principle, be described by the band theory of solids. Such a treatment naturally incorporates periodic boundary conditions and the long-range correlations of the neutron gas. The band-structure effect in the inner crust has been discussed in many works based on nonrelativistic density functional theory, where pairing correlations of nucleons were also taken into account \cite{chamel2007validity,chamel2012,yoshimura2024superfluid,almirante2026emergence}.} On the other hand, to easily describe the properties of the inner crust, the Wigner–Seitz (WS) cell approximation is commonly adopted to simplify the underlying periodic structure~\cite{negele1973neutron}. The nucleon–nucleon interactions for the neutron-rich nuclei in the inner crust are usually treated within nuclear density functional theory, such as the Skyrme–Hartree–Fock model~\cite{pearson2012skyrme}, the Gogny Hartree–Fock model~\cite{mondal20gogny}, and the relativistic mean-field (RMF) theory~\cite{ring1996relativistic,niksic2011relativistic,meng2006relativistic}. All these approaches have been shown to successfully describe the properties of nuclear matter and finite nuclei. 
			
			In general, the nuclei appearing in the inner crust contain a very large number of neutrons, so that many neutrons occupy the continuum states. Consequently, the choice of boundary conditions for these unbound states is essential for solving the corresponding Schr\"{o}dinger or Dirac equations. In the nonrelativistic framework, the first derivative of odd-parity wave functions vanishes at the cell radius, whereas even-parity wave functions vanish at the cell boundary~\cite{negele1973neutron}. However, for the Dirac equation, the boundary conditions at the cell radius are not uniquely defined and involve a certain degree of arbitrariness. As demonstrated by Cao \textit{et al.} in Ref.~\cite{Cao:2008cpl}, a direct extension of nonrelativistic boundary conditions to the relativistic regime violates the orthogonality of single-particle states. On the other hand, the semiclassical Thomas–Fermi (TF) approximation has been widely applied to calculate the equation of state (EOS) of the inner crust within RMF models~\cite{boguta1977thomas,shen1998relativistic,shchechilin2024unified}, where shell effects are completely neglected. This approach also enables computationally efficient large-scale investigations of neutron star matter.
			
		   Traditional techniques, such as the shooting method~\cite{horowitz1981self} and basis expansion approaches~\cite{gambhir1990relativistic,zhou2003spherical,li2007reflection,lu2014multidimensionally}, have been well established as robust tools for standard nuclear-structure problems. However, their application to the neutron-star inner crust faces significant challenges. As noted in Ref.~\cite{Zhang:2022spurious}, conventional methods can be highly sensitive to the choice of box size or require an extremely large basis space to accurately describe the weakly bound and continuum states that are characteristic of the inner crust.
		   
		   Alternatively, discretizing the Hamiltonian on a coordinate grid—known as the finite-difference method (FDM)~\cite{salomonson1989relativistic,fang2020solution,zhao2016spherical}—provides an efficient and accurate approach. Since this method does not rely on basis expansions, it is particularly well suited for describing states with extended spatial distributions. Despite these advantages, the direct application of FDM to the Dirac equation suffers from the so-called fermion-doubling problem, which is also well known in lattice quantum chromodynamics (QCD)~\cite{susskind1977lattice,stacey1982eliminating}. In particular, the standard central-difference discretization generates unphysical, rapidly oscillating spurious states that contaminate the physical spectrum. Although remedies such as the Wilson mass term are commonly employed in lattice QCD to suppress these modes, they inevitably introduce artificial modifications to the Hamiltonian.
			
			In this work, the asymmetric finite difference (AFD) method is adopted to overcome these numerical difficulties and to solve the radial Dirac equations for neutron-rich nuclei in WS cells. As proposed in Ref.~\cite{Zhang:2022spurious}, this method employs distinct finite-difference formulas for the large and small components of the Dirac spinor, depending on the sign of the quantum number $\kappa$. This prescription effectively eliminates spurious states for massive fermions without introducing artificial terms such as the Wilson mass, while preserving the hermiticity of the Hamiltonian matrix. As a result, the AFD method enables an accurate and fully quantum description of extremely neutron-rich nuclei in the inner crust of neutron stars.
			
			The paper is organized as follows. In Sec.~\ref{sec:theory}, we briefly outline the theoretical framework of the RMF theory and introduce the AFD method for solving the Dirac equations in the WS cell. Section~\ref{sec:results} presents the calculated results and discussion, focusing on the binding energies per nucleon, density distributions, and finally the neutron root-mean-square radii and neutron chemical potentials of the inner crust nuclei using TM1 family parameter sets. A summary is given in Sec.~\ref{sec:summary}.

			\section{Theoretical Framework}
			\label{sec:theory}
			
				\subsection{Relativistic Mean-Field Theory}
			
		In this work, the properties of nuclear systems are investigated within the RMF formalism. In this framework, the interaction between nucleons is described by the exchange of several mesons, including the isoscalar-scalar meson ($\sigma$), the isoscalar-vector meson ($\omega^\mu$), and the isovector-vector meson ($\rho^{a\mu}$). In addition, the electromagnetic interaction between protons is taken into account through the photon field ($A^\mu$). A cross-coupling term between the $\omega^\mu$ and $\rho^\mu$ mesons is included to adjust the density dependence of the nuclear symmetry energy and its slope~\cite{Bao:2014ese, fattoyev2010relativistic}. The Lagrangian density of the RMF model takes the form
		\begin{equation}
			\begin{split}
				\mathcal{L} = & \sum_{i=p,n}\bar{\psi}_{i}\left\{i\gamma_{\mu}\partial^{\mu}-(M+g_{\sigma}\sigma)-\gamma_{\mu}\left[g_{\omega}\omega^{\mu}+\frac{g_{\rho}}{2}\tau_{a}\rho^{a\mu}+\frac{e}{2}(1+\tau_{3})A^{\mu}\right]\right\}\psi_{i} \\
				& + \frac{1}{2}\partial_{\mu}\sigma\partial^{\mu}\sigma-\frac{1}{2}m_{\sigma}^{2}\sigma^{2}-\frac{1}{3}g_{2}\sigma^{3}-\frac{1}{4}g_{3}\sigma^{4}-\frac{1}{4}W_{\mu\nu}W^{\mu\nu}+\frac{1}{2}m_{\omega}^{2}\omega_{\mu}\omega^{\mu} \\
				& + \frac{1}{4}c_{3}(\omega_{\mu}\omega^{\mu})^{2}-\frac{1}{4}R_{\mu\nu}^{a}R^{a\mu\nu}+\frac{1}{2}m_{\rho}^{2}\rho_{\mu}^{a}\rho^{a\mu} \\
				& + \Lambda_{\mathrm{v}}(g_{\omega}^{2}\omega_{\mu}\omega^{\mu})(g_{\rho}^{2}\rho_{\mu}^{a}\rho^{a\mu})-\frac{1}{4}F_{\mu\nu}F^{\mu\nu}.
			\end{split}
			\label{eq:lagrangian}
		\end{equation}
		Here, $W^{\mu\nu}$, $R^{a\mu\nu}$, and $F^{\mu\nu}$ denote the antisymmetric field tensors associated with the $\omega^\mu$, $\rho^{a\mu}$, and $A^\mu$ fields, respectively. Within the mean-field approximation, the meson fields are treated as classical fields, and the corresponding field operators are replaced by their expectation values. As a result, the only non-vanishing classical fields are $\sigma = \langle\sigma\rangle$, $\omega = \langle\omega^0\rangle$, and $\rho = \langle\rho^{30}\rangle$. From the Lagrangian density given in Eq. \eqref{eq:lagrangian}, the equations of motion for the meson fields are obtained
		\begin{align}
			-\nabla^{2}\sigma+m_{\sigma}^{2}\sigma+g_{2}\sigma^{2}+g_{3}\sigma^{3}&=-g_{\sigma}\left(\rho_{p}^{s}+\rho_{n}^{s}\right), 
			\label{eq:sigma_eom} \\
			-\nabla^{2}\omega+m_{\omega}^{2}\omega+c_{3}\omega^{3}+2\Lambda_{\mathrm{v}}(g_{\omega}^{2}\omega)(g_{\rho}^{2}\rho^{2})&=g_{\omega}(\rho_{p}+\rho_{n}), 
			\label{eq:omega_eom} \\
			-\nabla^{2}\rho+m_{\rho}^{2}\rho+2\Lambda_{\mathrm{v}}(g_{\omega}^{2}\omega^{2})(g_{\rho}^{2}\rho)&=\frac{g_{\rho}}{2}\left(\rho_{p}-\rho_{n}\right), 
			\label{eq:rho_eom} \\
			-\nabla^2A&=e\rho_p.
			\label{eq:photon_eom}
		\end{align}
	      The source terms appearing in the right-hand sides of meson field equations are constructed from the nucleon wave functions, where $\rho_i^s$ denotes the scalar density and $\rho_i$ denotes the vector density for species $i=n,~p$. To evaluate these densities, the Dirac equation for the nucleons must first be solved in the present work. For a spherically symmetric nucleus, the single-particle nucleon wave function $\phi_\alpha$ is expressed as a two-component Dirac spinor
		\begin{equation}
			\phi_{\alpha}(\mathbf{r})=\frac{1}{r}\begin{pmatrix}\mathrm{i}G_{n\kappa}(r)\Phi_{\kappa m}(\hat{\mathbf{r}})\\-F_{n\kappa}(r)\Phi_{-\kappa m}(\hat{\mathbf{r}})\end{pmatrix}.
			\label{eq:nucleon_wf}
		\end{equation}
		In this expression, $G_{n\kappa}(r)$ and $F_{n\kappa}(r)$ denote the radial wave functions of the large and small components, respectively. $\Phi_{\kappa m}(\hat{\mathbf{r}})$ represents the spin–spherical harmonics. The single-particle state $\alpha$ is characterized by the set of quantum numbers ${n, \kappa, m}$, where $n$ is the radial quantum number, $\kappa$ is the relativistic angular-momentum quantum number defined as $\kappa = \pm (j+1/2)$, and $m$ is the projection of the total angular momentum $j$ on the $z$-axis. The wave function is normalized to unity, and its radial dependence is governed by the following set of coupled equations
			\begin{align}
			\frac{\mathrm{d}}{\mathrm{d}r}G_{\alpha}(r)+\frac{\kappa}{r}G_{\alpha}(r) &= M^{*}F_{\alpha}(r)-\left[g_{\omega}\omega+\frac{g_{\rho}}{2}\tau_{3}\rho+\frac{e}{2}(1+\tau_{3})A\right]F_{\alpha}(r)+E_{\alpha}F_{\alpha}(r), \label{eq:dirac_G} \\
			\frac{\mathrm{d}}{\mathrm{d}r}F_{\alpha}(r)-\frac{\kappa}{r}F_{\alpha}(r) &= M^{*}G_{\alpha}(r)+\left[g_{\omega}\omega+\frac{g_{\rho}}{2}\tau_{3}\rho+\frac{e}{2}(1+\tau_{3})A\right]G_{\alpha}(r)-E_{\alpha}G_{\alpha}(r), \label{eq:dirac_F}
		\end{align}
		where $M^* = M + g_\sigma \sigma$ denotes the effective nucleon mass. Once the nucleon wave functions are determined, the source terms entering the meson field equations can be evaluated. The scalar and vector densities are then constructed from the wave-function components as follows
		\begin{align}
			\rho_i^s(r) &=\left[ \sum_{\alpha}^{\text{occ}} w_\alpha \frac{2j_\alpha + 1}{4\pi r^2} \left( |G_\alpha(r)|^2 - |F_\alpha(r)|^2 \right)\right]_i, \label{eq:scalar_density} \\
			\rho_i(r) &= \left[\sum_{\alpha}^{\text{occ}} w_\alpha \frac{2j_\alpha + 1}{4\pi r^2} \left( |G_\alpha(r)|^2 + |F_\alpha(r)|^2 \right)\right]_i. \label{eq:vector_density}
		\end{align}
		Here, the summation runs over all occupied states, and $w_\alpha$ denotes the corresponding occupation probability. {In the inner crust of neutron stars, dripped neutrons are generally expected to form spin-singlet Cooper pairs through the attractive interaction in the $^1S_0$ channel. These pairing correlations can have important impacts on a variety of neutron-star phenomena, including the microscopic properties of the crust, its thermal behavior, and its response to external perturbations.} However, in the present work, pairing correlations are not taken into account. Consequently, the occupation probability $w_{\alpha}$ is determined using the uniform filling approximation. 
		It is evident that the meson field equations and the Dirac equations for nucleons are mutually coupled. The mean-field potentials determine the nucleon wave functions, which in turn generate the source densities for the meson fields. Consequently, this coupled system must be solved self-consistently through an iterative procedure. In the present work, the iteration is continued until both the potentials and the densities converge within a prescribed numerical accuracy.
		
	\subsection{Asymmetric Finite Difference method}
		To solve the Dirac equation with the AFD method, the calculation is performed within a finite box of radius $R_c$. This box radius is discretized into $N$ equidistant grid points, $r_i = i \times h$ (for $i=1, ..., N$) with a step size $h = R_c / N$, allowing the radial wave functions $G(r)$ and $F(r)$ to be represented as $N$-dimensional vectors for a fixed $\kappa$. In the AFD method, the first-order derivative $\mathrm{d}/\mathrm{d}r$ is approximated using asymmetric difference formulas. In this work, we specifically employ the five-point approximation. For a general function $f(r)$, the forward difference formula at the grid point $r_i$ is given by
		\begin{equation}
			\frac{\mathrm{d}f}{\mathrm{d}r}\bigg|_i = \frac{-25f_{i} + 48f_{i+1} -  36f_{i+2}+16f_{i+3}-3f_{i+4}}{12h}.
		\end{equation}			
		The corresponding backward difference formula is			 
		\begin{equation}
			\frac{\mathrm{d}f}{\mathrm{d}r}\bigg|_i = \frac{25f_{i} - 48f_{i-1} +  36f_{i-2}-16f_{i-3}+3f_{i-4}}{12h}.
		\end{equation}
		This discretization transforms the derivative operator into a sparse matrix. The backward difference formula, for example, is represented by a lower-banded matrix 
		\begin{equation}
			\frac{\mathrm{d}}{\mathrm{d}r}=
			\begin{pmatrix}
				\frac{25}{12h}&&&&&&\\
				-\frac{48}{12h}&\frac{25}{12h}&&&&&\\
				\vdots&&\ddots&&&&\\
				\vdots&&&\ddots&&&\\
				\vdots&&&&\ddots&&\\
				\vdots&&&&&\ddots&\\
				0&\cdots&\frac{3}{12h}&-\frac{16}{12h}&\frac{36}{12h}&-\frac{48}{12h}&\frac{25}{12h}
			\end{pmatrix}.
		\end{equation}		 
		The forward difference formula similarly leads to an upper-banded matrix structure. By applying the discretization scheme described above to the radial Dirac equations \eqref{eq:dirac_G} and \eqref{eq:dirac_F}, the original differential eigenvalue problem is converted into a $2N \times 2N$ matrix eigenvalue problem. This problem can be written in the following matrix form
		\begin{equation}
			\left(
			\begin{array}{c|c}
				\mathbf{A} & \mathbf{B_1} \\
				\hline
				\mathbf{B_2} & \mathbf{C}
			\end{array}
			\right)
			\begin{pmatrix} \mathbf{G} \\ \mathbf{F} \end{pmatrix}
			= E
			\begin{pmatrix} \mathbf{G} \\ \mathbf{F} \end{pmatrix}.
		\end{equation}
	In this formulation, the diagonal blocks $\mathbf{A}$ and $\mathbf{C}$ contain the potential and mass terms, whereas the off-diagonal blocks $\mathbf{B_1}$ and $\mathbf{B_2}$ incorporate the derivative terms as well as the centrifugal terms proportional to $\kappa/r$. 
	The boundary condition is that the wave functions are assumed to be zero at $r = 0$ and outside the box $r > R_c$~\cite{Zhang:2022spurious}. 
	To eliminate spurious states while preserving the hermiticity of the Hamiltonian matrix, the choice of the finite-difference scheme is crucial. The finite difference formulas are selected according to the sign of $\kappa$. For $\kappa > 0$, the large component $\mathbf{G}$ is discretized using a backward difference form, while the small component $\mathbf{F}$ is treated with a forward difference one, and the assignment is reversed for $\kappa < 0$. This distinction underscores the necessity of treating states with different signs of $\kappa$ separately in numerical implementations. The present approach follows the methodology of Ref.~\cite{Zhang:2022spurious}. A similar consideration arises in the treatment of boundary conditions, where Cao \textit{et al.}~\cite{Cao:2008cpl} demonstrated that strict preservation of the orthogonality of single-particle states can be achieved using a scheme that depends on the sign of $\kappa$, specifically by imposing a vanishing boundary condition on the appropriate component.
	
	\section{Numerical results and discussions}	\label{sec:results}
		In the present work, the properties of extremely neutron-rich nuclear systems are investigated within the WS cell approximation. It should be noted that the present setup does not aim at a fully realistic description of the neutron-star inner crust, since the electron gas and the neutron gas under $\beta$-equilibrium conditions are not explicitly taken into account. To systematically explore the influence of the density dependence of the symmetry energy on inner-crust structures, a family of parameter sets based on the TM1 interaction is adopted~\cite{Bao:2014ese}, whose slope of symmetry energy, $L$, lies in the range $40\sim 110$ MeV. {The symmetry-energy slope $L$ determines the density dependence of the symmetry energy and is strongly correlated with the neutron-skin thickness of finite nuclei and the radii of neutron stars. However, the present experimental and observational constraints still involve large uncertainties. The PREX-II experiment favors a relatively large value, greater than $100$ MeV, whereas astrophysical observations of neutron stars and gravitational waves suggest a value around $40$--$80$ MeV. Therefore, in the present work, we choose $L$ in the range of $40$--$110$ MeV \cite{li2026beyond}.}
		
		In this scheme, the couplings of the isoscalar-scalar and isoscalar-vector mesons are kept fixed at their original TM1 values, thereby preserving the saturation properties of symmetric nuclear matter. The isovector meson couplings, $g_\rho$ and $\Lambda_{\mathrm{v}}$, are adjusted to generate different values of the symmetry-energy slope $L$ at nuclear saturation density, while the symmetry energy is constrained to be identical at a sub-saturation reference density of $n_{\text{fix}} = 0.11 \text{ fm}^{-3}$. Specifically, calculations are performed using parameter sets with $L = 40$, $60$, and $80$ MeV, together with the original TM1 parameter set corresponding to $L = 110.8$ MeV.
		
		In addition, to examine the effects of the effective nucleon mass, the recently developed TM1m* parameter set is also employed~\cite{Li:2025nmf}, since the EOS with larger effective nucleon mass makes supernova explosions more likely to occur.  This interaction features a larger effective mass ($M^*/M \approx 0.79$) than the standard TM1 value ($M^*/M \approx 0.63$) and has been adjusted to better reproduce the ground-state properties of finite nuclei. The TM1m* interaction shares the same symmetry energy slope ($L = 40$ MeV) as the corresponding model in the TM1-based series discussed above. The detailed coupling constants of all parameter sets used in this work have been listed in Table~\ref{tab:parameters}, and the remaining parameters are the same as the original TM1 set \cite{sugahara1994relativistic}.
	\begin{table}[htbp]
		\centering
		\caption{The coupling constants of the TM1 family parameter sets used in this work. The masses $m_\sigma$ are in MeV, and the nonlinear coupling $g_2$ is in fm$^{-1}$. Other parameters are dimensionless. A hyphen indicates that the corresponding parameter is the same as that in the original TM1 set.}
		\label{tab:parameters} 
		\setlength{\tabcolsep}{12pt}  
		\begin{tabular}{lccccc}
			\toprule
			& TM1&$L$=40 & $L$=60 & $L$=80 & TM1m* \\
			\midrule
			$m_\sigma$   & 511.198              & -  & - & - & 463.680 \\
			$g_\sigma$     & 10.0289           & -   & - & - & 7.1545  \\
			$g_\omega$    & 12.6139            & -   & - & - & 8.59221  \\
			$g_\rho$           & 9.2644            & 13.9714  & 11.2610  & 10.1484   & 11.5216   \\
			$g_2$                & $-7.2325$     & -& - & - & $-7.8668$ \\
			$g_3$                & 0.6183            & -  & -  & - & $40.55692$ \\
			$c_3$                & 71.3075          & -  & - & - & 0.0       \\
			$\Lambda_{\text{v}}$  & 0.0   & 0.0429  & 0.0248  & 0.0128     & 0.0947       \\
			\bottomrule
		\end{tabular}
	\end{table}
	
	In this work, the nine WS cell configurations with different nucleon mass numbers, $A$, for Zr and Sn listed in Ref.~\cite{Cao:2008cpl} are adopted. These configurations correspond to baryon densities ranging from $\rho = 0.0016 \rho_0$ to $0.12 \rho_0$, where $\rho_0 = 0.17 \text{ fm}^{-3}$. {In the original many-body study of the neutron-star inner crust~\cite{negele1973neutron}, it was shown that, at a given baryon density in the inner crust, local minima of the nuclear binding energy arise from shell closures at 40 and 50 protons when pairing correlations are neglected. The ground-state solutions obtained in that work for densities from $4.6\times10^{11}$ g/cm$^3$ to $3.4\times10^{13}$ g/cm$^3$ correspond to nuclei ranging from $^{180}$Zr to $^{1800}$Sn, as listed in Table II. Subsequently, many studies investigated these nuclei in order to discuss the structure of the neutron-star inner crust, such as Refs.\cite{sandulescu2004nuclear,Cao:2008cpl,grill2011cluster}. Therefore, we also adopt these isotopes here for comparison.} The detailed parameters for each layer, including the proton number $Z$, neutron number $N$, and the cell radius $R_c$, are summarized in Table~\ref{tab:layers}.
	\begin{table}[htbp]
		\centering
		\caption{Adopted WS cell configurations for the nine inner crust layers ($\rho_0 = 0.17$ fm$^{-3}$).}
		\label{tab:layers}
		\begin{tabular}{ccccc}
			\toprule
			Layer &~ $\rho/\rho_0$ ~& ~$Z$ ~& ~$N$ ~&~ $R_c$ (fm) ~\\
			\midrule
			1 & 0.0016 & 40 & 140 & 53.6 \\
			2 & 0.0024 & 40 & 160 & 49.2 \\
			3 & 0.0035 & 40 & 210 & 46.3 \\
			4 & 0.0052 & 40 & 280 & 44.3 \\
			5 & 0.0093 & 40 & 460 & 42.2 \\
			6 & 0.022  & 50 & 900 & 39.3 \\
			7 & 0.034  & 50 & 1050 & 35.7 \\
			8 & 0.052  & 50 & 1300 & 33.1 \\
			9 & 0.12   & 50 & 1750 & 27.6 \\
			\bottomrule
		\end{tabular}
	\end{table}	 
			 
		Based on the WS cell configurations detailed in Table~\ref{tab:layers}, we proceed to calculate the binding energy per nucleon $E/A$ for these neutron-rich nuclei appearing in the inner crust. These calculations are performed to evaluate the energetic properties of the system under extreme neutron-rich conditions. {For comparison, we also perform calculations within the TF approximation. It follows the self-consistent relativistic TF approach, whose details were shown in Ref. \cite{Li:2025nmf}. In this approximation, the shell effect is neglected and the nucleons are treated as locally uniform Fermi gases. The local scalar and vector densities are determined from the local Fermi momenta, while the neutron and proton chemical potentials are spatially constant in the WS cell. For a fixed cell radius $R_c$ and given proton and neutron numbers, we start from trial mean fields (or equivalently trial density distributions), determine the corresponding chemical potentials, calculate the local densities, and solve the meson-field equations Eqs. (\ref{eq:sigma_eom}-\ref{eq:photon_eom})  self-consistently. The iteration is repeated until the densities, chemical potentials, and mean fields converge. Therefore, the TF results presented in this work correspond to the same prescribed WS cells as those adopted in the AFD calculations and serve as semiclassical reference results for comparison.} The values of $E/A$ for the nine inner crust layers, employing the TM1 family parameter sets with different symmetry energy slopes $L$ and effective masses, are summarized in Table~\ref{tab:perE}.
			\begin{table}[htbp]
			\caption{The binding energy per nucleon ($E/A$ in MeV) for the nine inner crust layers calculated using different parameter sets ($L=40, 60, 80, 110.8$ MeV and TM1m*) within the TF approximation and the AFD method.}
			\centering
			%
				\begin{tabular}{c c c c c c c c c c c}
				\hline\hline
				\multirow{2}{*}{Layer}
				& \multicolumn{2}{c}{\makecell{$L=40$}}
				& \multicolumn{2}{c}{\makecell{$L=60$}}
				& \multicolumn{2}{c}{\makecell{$L=80$}}
				& \multicolumn{2}{c}{\makecell{TM1\\($L=110.8$)}}
				& \multicolumn{2}{c}{\makecell{TM1m*\\($L=40$)}} \\
				%
			& ~~TF~~ & ~~AFD~~ & ~~TF~~ & ~~AFD~~ & ~~TF~~ & ~~AFD~~ & ~~TF~~ & ~~AFD~~ & ~~TF~~ & ~~AFD~~ \\
				\hline
				1 & $-4.940$ & $-4.958$ & $-5.053$ & $-5.040$ & $-5.113$ & $-5.090$ & $-5.169$ & $-5.168$ & $-5.363$ & $-5.062$ \\
				2 & $-4.389$ & $-4.391$ & $-4.499$ & $-4.476$ & $-4.556$ & $-4.538$ & $-4.611$ & $-4.608$ &   $-4.775$ & $-4.510$ \\
				3 & $-3.365$ & $-3.334$ & $-3.470$ & $-3.428$ & $-3.523$ & $-3.500$ & $-3.573$ & $-3.558$ & $-3.685$ & $-3.469$ \\
				4 & $-2.398$ & $-2.310$ & $-2.508$ & $-2.457$ & $-2.560$ & $-2.520$ & $-2.610$ & $-2.572$ & $-2.667$ & $-2.481$ \\
				5 & $-0.956$& $-0.850$ & $-1.108$ & $-1.022$ & $-1.173$ & $-1.091$ & $-1.227$ & $-1.126$ & $-1.183$ & $-1.012$ \\
				6 & 0.512 & 0.671 & 0.190 & 0.334 & 0.069 & 0.201 & $-0.025$ & 0.099 & 0.216 & 0.416 \\
				7 & 1.264 & 1.484 & 0.773 & 0.955 & 0.592 & 0.757 & 0.453 & 0.608 & 0.855 & 1.090 \\
				8 & 2.265 & 2.557 & 1.501 & 1.740 & 1.224 & 1.433 & 1.016 & 1.200 & 1.662 & 1.953 \\
				9 & 4.714 & 5.226 & 3.046 & 3.467 & 2.428 & 2.792 & 1.968 & 2.277 & 3.387 & 3.828 \\
				\hline\hline
			\end{tabular}
			\label{tab:perE}
		\end{table}
		
		When examining the TM1 family of parameter sets, a systematic trend is observed where the binding energy per nucleon decreases as the symmetry energy slope $L$ increases.  This suggests that higher values of $L$ correspond to a lower total energy for the specified inner crust configurations, since the slope of symmetry energy plays a negligible role below the saturation density $\rho_0$, compared to those above $\rho_0$. The smaller $L$ will generate a bigger symmetry energy for the density region of finite nuclei. It means that the interaction in neutron-rich nuclei becomes more repulsive, reducing the binding energy.
		
	  {It should be emphasized that the present discussion on the $L$ dependence of $E/A$ is made for the fixed WS cell configurations adopted in Table II. In realistic inner-crust calculations, however, the $\beta$-equilibrium condition should be imposed and the energetically optimal configuration should be determined at each baryon density. In that case, the optimum proton fraction, as well as the corresponding neutron and proton numbers, would generally depend on $L$. Therefore, the influence of $L$ in the inner crust is reflected not only in the energy variation of a given configuration, but also in the possible change of the equilibrium composition.} In comparing two interactions with the same slope of $L=40$ MeV, the TM1m* set consistently yields lower $E/A$ values than the corresponding TM1-based set. This discrepancy is particularly significant in layers with higher nucleon densities, indicating that the larger effective mass in the TM1m* set plays a vital role in reducing the system's energy. 
		
		{This discrepancy is particularly significant in layers with higher nucleon densities. It should be noted that the lower $E/A$ obtained with the TM1m* interaction should not be attributed solely to the kinetic-energy reduction associated with a larger effective mass. Since TM1m* differs from the TM1-based $L = 40$ set also in the self-consistent mean-field parameters, the reduction of the total energy should be understood as a combined effect of the larger effective mass and the corresponding rearrangement of the scalar and vector mean fields, which modifies both the kinetic and potential-energy contributions.}
		
		Regarding the numerical methods employed, results obtained from the TF approximation are generally lower than those derived from the AFD method. For the TM1-based parameter sets (excluding TM1m*), the results from TF and AFD are quite similar in the lower density region (Layers 1–5), whose binding energies are less than zero. However, the differences become more pronounced for the TM1m* set, which has been shown in Ref. \cite{Li:2025nmf}, as well as in higher density layers. This variance is typically attributed to the inclusion of quantum shell effects and gradient terms in the AFD method, which are often averaged out or neglected in the semiclassical approximation.
		
			\begin{figure}[htbp]
			\centering
			\includegraphics[width=0.6\linewidth]{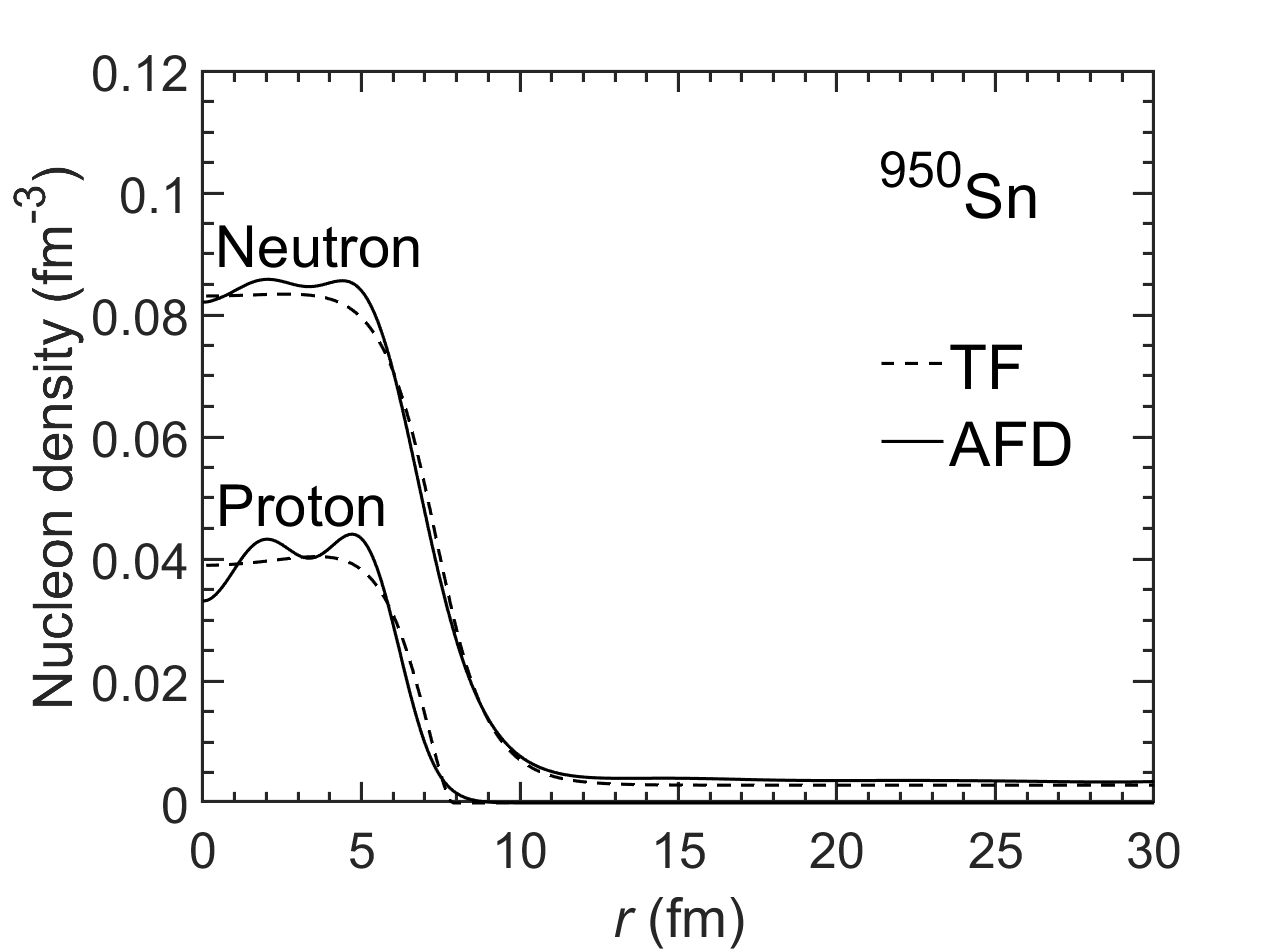}
			\caption{Proton and neutron density distributions of $^{950}$Sn calculated using the TM1 interaction ($L=110.8$ MeV). The solid lines represent the results from the AFD approach, while the dashed lines correspond to the TF approximation.}
			\label{fig:Sn950}
		\end{figure}
		
		In Fig.~\ref{fig:Sn950}, we illustrate the proton and neutron density distributions of $^{950}$Sn, which is an extremely neutron-rich nucleus, calculated using the original TM1 interaction ($L=110.8$ MeV) within both the AFD method and TF approximation. The solid lines depict the results obtained from the AFD method, while the dashed lines represent those derived from the TF approximation.
		It is clear that the nucleon densities produced by the AFD method display notable fluctuations, particularly in the central region of the nucleus ($r < 7$ fm). These oscillations in the densities obviously arise from quantum shell effects, reflecting the discrete nature of the single-particle wave functions~\cite{Li:2025nmf}. In contrast, the TF approximation yields smooth and flat density distributions. Additionally, at larger distances ($r > 10$ fm), a finite and constant neutron density is observed, indicating the presence of a surrounding neutron gas in the inner crust environment. In contrast, the proton density rapidly reduces to zero around $r=9$ fm due to the small $Z$. 
		
		\begin{figure}[htbp]
			\includegraphics[width=1.0\linewidth]{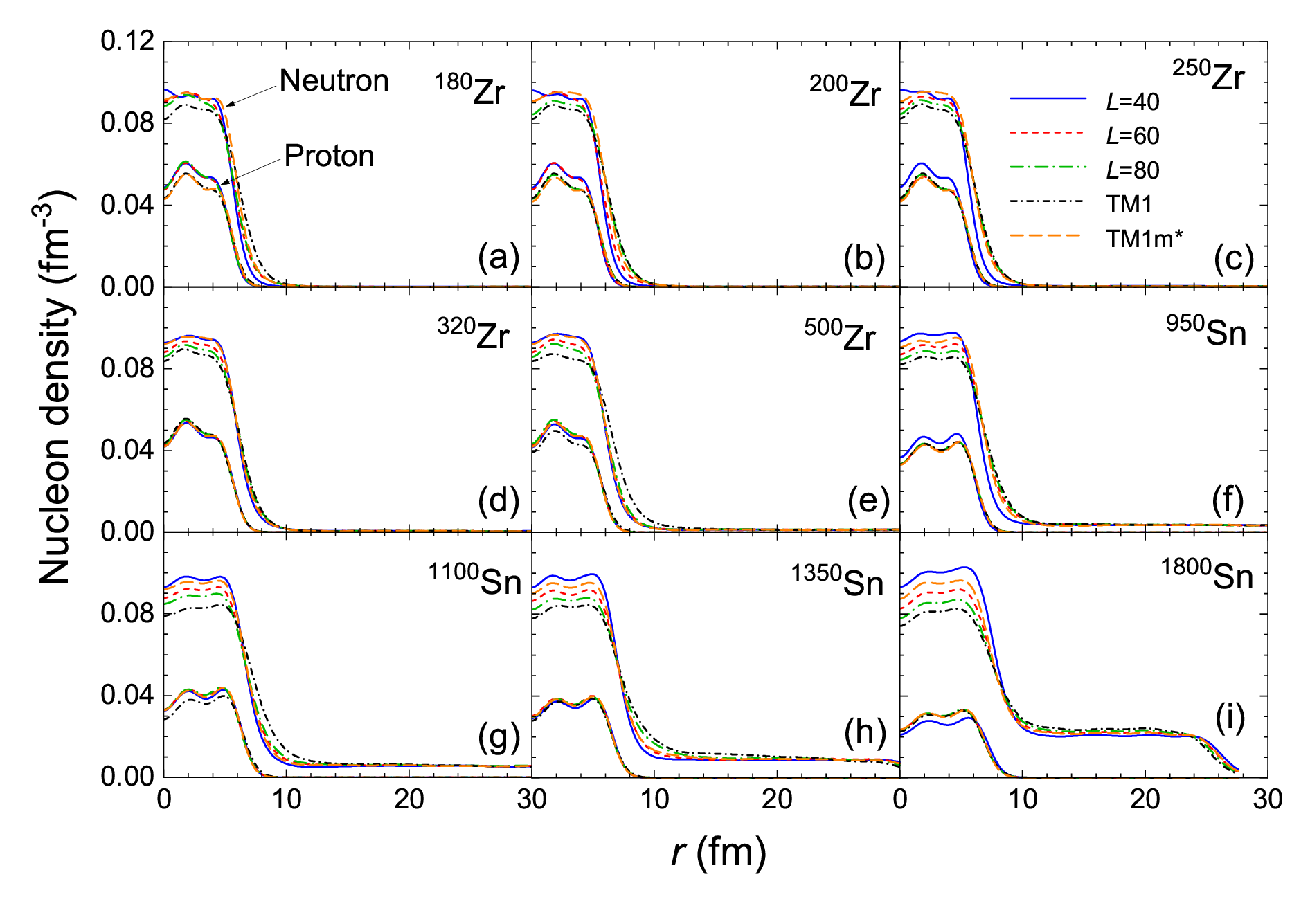}
			\caption{Proton and neutron density distributions in the WS cells for the nine inner crust layers, where panels (a)-(i) correspond to Layers 1-9. The results are calculated using the AFD method with different parameter sets, including $L=40, 60, 80, 110.8$ MeV from the TM1 family and the TM1m* interaction.}
			\label{fig:afdpn}
		\end{figure}
		In Fig.~\ref{fig:afdpn}, the nucleon density distributions for the nine layers of the inner crust are presented, calculated using different parameter sets.
		Initially focusing on the neutron distributions within the TM1-based family, which keeps the identical isoscalar properties, we observe a systematic dependence of density distributions on the symmetry energy slope $L$. In the central region of the WS cell, the neutron density exhibits an inverse correlation with $L$. The parameter set with $L=40$ MeV displays the highest central density, whereas the original TM1 set ($L=110.8$ MeV) yields the lowest density. However, moving to the nuclear surface region, this trend reverses, with parameter sets featuring larger $L$ generally corresponding to higher neutron densities. As discussed in Ref.~\cite{Bao:2014ese}, this redistribution occurs because a larger $L$ yields a larger symmetry energy in the dense central region, which favors reducing the central neutron density and enhancing the distribution at the surface. However, in the outer tail region, the parameter sets with smaller $L$ eventually sustain a higher neutron density than those with larger  $L$. While this effect is subtle and almost imperceptible in the lower-density layers [Figs.~\ref{fig:afdpn}(a)--\ref{fig:afdpn}(e)], it becomes distinctively clearly shown in the boundary drop-off region of Fig.~\ref{fig:afdpn}(i). Regarding proton distributions, they show relatively less sensitivity to variations in $L$. However, minor differences among the curves are still apparent, since the protons and neutrons are coupled with each other in the self-consistent mean field through the isovector meson. The proton densities with small $L$ will be suppressed as the neutron number increases.
		
		In examining the impact of the effective nucleon mass, we compare the TM1m* set with the TM1-based $L=40$ MeV set, since both of them have the same slope of symmetry energy. The first three panels Figs.~\ref{fig:afdpn}(a)-(c) for Layers 1–3 reveal a crossover behavior in the central region, where the TM1-based $L=40$ MeV set shows a higher peak at the center but its density decreases more rapidly with increasing radius. As a result, the TM1m* density eventually exceeds that of the $L=40$ MeV set in the outer part of the central region. In Figs.~\ref{fig:afdpn}(d)-(e) for Layers 4 and 5, the density profiles of both parameter sets become quite similar in the central area. 	
		
		However, a distinct trend emerges in the Figs.~\ref{fig:afdpn}(f)-(i) for Layers 6–9 for Sn isotopes, which correspond to extremely neutron-rich conditions. In these layers, the TM1m* profile consistently reveals a lower central density compared to the TM1-based $L=40$ set throughout the central region, and its curve generally resides between those of $L=40$ and $L=60$. This observation suggests that, for a fixed symmetry energy slope, an increase in effective nucleon mass tends to diminish the nucleon density in the central region, producing an effect comparable to that of increasing the slope $L$ while maintaining the effective mass. The larger effective mass will generate a smaller spin-orbital coupling force.
		
		Furthermore, it can be found that the density distributions of neutrons for $^{1800}$Sn close to the boundary have sharp decreases. This is due to the boundary condition mentioned in Sec. II B.  
		This decrease is independent of the boundary conditions shown in Ref. \cite{Cao:2008cpl}, since it was also observed in Fig. 1 of that paper.
			\begin{table}[htbp]
			\centering
			\caption{Neutron root-mean-square radii $R_n$ (in fm) for the nine inner crust layers calculated using the TF approximation and the AFD method with different parameter sets.}
			\begin{tabular}{c c c c c c c c c c c}
				\hline\hline
				\multirow{2}{*}{Layer}
				& \multicolumn{2}{c}{\makecell{$L=40$}}
				& \multicolumn{2}{c}{\makecell{$L=60$}}
				& \multicolumn{2}{c}{\makecell{$L=80$}}
				& \multicolumn{2}{c}{\makecell{TM1\\($L=110.8$)}}
				& \multicolumn{2}{c}{\makecell{TM1m*\\($L=40$)}} \\
				%
				& ~~TF~~ & ~~AFD~~ & ~~TF~~ & ~~AFD~~ & ~~TF~~ & ~~AFD~~ & ~~TF~~ & ~~AFD~~ & ~~TF~~ & ~~AFD~~ \\
				\hline
				1 & 26.278 & 24.591 & 24.933 & 22.344 & 24.031 & 22.278 & 23.008 & 17.622 & 25.394 & 19.413 \\
				2 & 26.176 & 24.894 & 25.157 & 23.182 & 24.475 & 19.916 & 23.698 & 19.890 & 25.508 & 20.005 \\
				3 & 27.584 & 25.956 & 26.899 & 23.197 & 26.435 & 23.133 & 25.900 & 23.089 & 27.139 & 23.252 \\
				4 & 28.465 & 25.712 & 27.986 & 25.216 & 27.649 & 25.140 & 27.255 & 25.084 & 28.155 & 25.279 \\
				5 & 29.277 & 27.645 & 28.999 & 27.085 & 28.787 & 26.991 & 28.525 & 25.860 & 29.099 & 27.521 \\
				6 & 28.328 & 27.113 & 28.191 & 26.387 & 28.052 & 26.247 & 27.854 & 26.124 & 28.238 & 26.609 \\
				7 & 25.946 & 24.757 & 25.867 & 24.325 & 25.754 & 24.143 & 25.576 & 23.453  & 25.894 & 24.419 \\
				8 & 24.296 & 23.176 & 24.289 & 22.901 & 24.211 & 22.439 & 24.068 & 21.867 & 24.294 & 22.790 \\
				9 & 20.393 & 19.529 & 20.581 & 19.424 & 20.597 & 19.289 & 20.572 & 19.119 & 20.547 & 19.448 \\
				\hline\hline
			\end{tabular}
			\label{tab:R_n}
		\end{table}
		
		The calculated neutron root-mean-square radii, $R_n$, for the nine layers of the inner crust are presented in Table~\ref{tab:R_n}.
		The dependence of $R_n$ on $L$ within the TM1 family reveals a notable trend where the neutron radius $R_n$ decreases as $L$ increases. This pattern is particularly pronounced in the layers with lower densities (e.g., Layers 1–6), highlighting the influence of the symmetry energy on the spatial distribution of neutrons. This indicates that despite the expansion in the surface region for larger $L$, the higher neutron density in the outer tail region for the low-$L$ sets (as discussed in Fig.~\ref{fig:afdpn}) plays a dominant role in determining the total radius. 
		
		In addition, the difference in $R_n$ between large and small $L$ is more obvious in the AFD method than in the TF approximation. This is because the neutron gas density from the AFD method decays to zero close to the boundary. Notably, the magnitude of this discrepancy depends heavily on the layer's average density. The difference between the TF and AFD results is most significant in the low-density layers (e.g., Layer 1) and gradually diminishes as the layer number increases. By Layer 9, where the density is highest, the results from the two methods become much closer, suggesting that the quantum shell effects captured by the AFD method are more pronounced in dilute environments.
		
		In examining the impact of the effective nucleon mass, we focus on a comparison between the TM1m* parameter set and the TM1-based $L=40$ configuration. The results indicate that the TM1m* interaction produces smaller neutron radii than the TM1-based $L=40$ set. This trend suggests that an increased effective mass is associated with a more concentrated neutron distribution within the WS cell. 
		
		\begin{table}[htbp]
			\centering
			\caption{Neutron chemical potentials $\mu_n$ (in $\mathrm{MeV}$) for the nine inner crust layers calculated using the TF approximation and the AFD method with different parameter sets.}
			%
				\begin{tabular}{c c c c c c c c c c c}
				\hline\hline
				\multirow{2}{*}{Layer}
				& \multicolumn{2}{c}{\makecell{$L=40$}}
				& \multicolumn{2}{c}{\makecell{$L=60$}}
				& \multicolumn{2}{c}{\makecell{$L=80$}}
				& \multicolumn{2}{c}{\makecell{TM1\\($L=110.8$)}}
				& \multicolumn{2}{c}{\makecell{TM1m*\\($L=40$)}} \\
				%
				& ~~TF~~ & ~~AFD~~ & ~~TF~~ & ~~AFD~~ & ~~TF~~ & ~~AFD~~ & ~~TF~~ & ~~AFD~~ & ~~TF~~ & ~~AFD~~ \\
				\hline
				1 & 0.355 & 0.447 & 0.314 & 0.318 & 0.293 & 0.309 & 0.273 & 0.297 & 0.328 & 0.318 \\
				2 & 0.506 & 0.652 & 0.450 & 0.503 & 0.424 & 0.435 & 0.400 & 0.428 & 0.469 & 0.453 \\
				3 & 0.781 & 0.940 & 0.693 & 0.723 & 0.658 & 0.706 & 0.627 & 0.694 & 0.723 & 0.740 \\
				4 & 1.117 & 1.256 & 0.981 & 1.131 & 0.929 & 1.085 & 0.887 & 1.051 & 1.027 & 1.176 \\
				5 & 1.827 & 1.963 & 1.553 & 1.692 & 1.453 & 1.610 & 1.377 & 1.471 & 1.642 & 1.779 \\
				6 & 3.250 & 3.550 & 2.568 & 2.752 & 2.326 & 2.524 & 2.148 & 2.360 & 2.773 & 2.964 \\
				7 & 4.269 & 4.656 & 3.209 & 3.528 & 2.833 & 3.102 & 2.558 & 2.864 & 3.506 & 3.809 \\
				8 & 5.622 &  6.202 & 3.987 & 4.354 & 3.403 & 3.661 & 2.978 & 3.280 & 4.402 & 4.639 \\
				9 & 9.416 & 10.003 & 6.106 & 6.885 & 4.843 & 5.703 & 3.900 & 4.703 & 6.726 & 7.232 \\
				\hline\hline
			\end{tabular}
			\label{tab:miun}
		\end{table}
		
		Table~\ref{tab:miun} presents the neutron chemical potentials $\mu_n$ for the nine inner crust layers calculated with different parameter sets, which will be very important for the $\beta$-equilibrium conditions. It is noteworthy that $\mu_n$ demonstrates a significant decrease with an increasing symmetry energy slope $L$. For instance, in Layer 9, the chemical potentials calculated with the $L=40$ MeV parameter set are more than twice those obtained from the original TM1 set, both for the TF approximation and AFD methods. This means that more high-angular-momentum orbitals will be occupied at smaller $L$. In terms of the effective nucleon mass, the TM1m* interaction yields systematically lower chemical potentials when compared to the $L=40$ MeV set, despite both having the same slope. 
		This reduction is attributed to the larger effective mass in the TM1m* interaction, which suppresses the kinetic energy contribution to the Fermi energy.

	   {When comparing the two numerical methods, the neutron chemical potentials obtained from the AFD method are generally larger than those derived from the TF approximation.} This indicates that quantum shell effects typically enhance the chemical potential in these inner crust environments. However, a slight reversal is observed for the TM1m* in the dilute Layers 1 and 2, where the values from TF approximation marginally exceed the AFD ones. This anomalous behaviour aligns with the findings in Ref.~\cite{Li:2025nmf}, which reported that parameter sets with large effective masses (such as TM1m*) can exhibit an opposite trend in the deviations between semiclassical and quantum calculations compared to other models like the original TM1 and the TM1-based $L=40$. As suggested in Ref.~\cite{Li:2025nmf}, this difference is largely attributed to specific nuclear matter properties, particularly the effective mass. Interestingly, this distinctive feature appears to be specific to the low-density regime, as the TM1m* results represent a return to the general trend observed in the denser layers.
			
			\section{Summary}
			\label{sec:summary}
			
		In this work, we systematically explore the ground-state properties of extremely neutron-rich nuclear systems in the inner crust of neutron stars, employing the RMF formalism within the Wigner-Seitz approximation. By adopting a series of TM1-family parameter sets distinguished by varying symmetry energy slopes $L$ and effective  nucleon masses, we analyzed the effects of the symmetry-energy slope $L$ and the effective nucleon mass $M^*$ on the inner crust properties. To address the spurious-state problem in the numerical procedure, we implement the AFD method to solve the radial Dirac equations, and compare the results with those derived from the TF approximation.
		
		{The investigation demonstrates that, in contrast to the smooth density profiles yielded by the TF approximation, the AFD method captures significant density fluctuations in the central region and leads to noticeable differences in the neutron chemical potentials. These shell effects may have important influences on the microscopic structure, equilibrium composition, superfluid-related properties, and transport properties of representative Wigner–Seitz cells in the inner crust, particularly in the low-density region. This indicates that quantum shell effects are non-negligible for the microscopic properties of the neutron-star inner crust.}
		
		On this basis, the study highlights the crucial influence of the nuclear equation of state parameters on the inner crust structure. It is found that the symmetry energy slope $L$ and the effective nucleon mass exhibit similar critical roles in regulating system properties. Specifically, an increase in either the slope $L$ or the effective nucleon mass leads to a stronger binding of neutrons, resulting in a more compact neutron distribution within the cell and a significant reduction in the binding energy. These findings elucidate that under extremely neutron-rich conditions, the density dependence of the symmetry energy and the effective nucleon mass constitute the core physical mechanisms determining the microscopic structure and energetic characteristics of the inner crust. 
		
		The $\beta$-equilibrium conditions will be considered in future work to construct the inner-crust EOS. {Furthermore, periodic boundary conditions in the Dirac equation will be treated using the AFD method in order to investigate the band-structure effect in the inner crust. We also note that neutrons in the inner crust are expected to exhibit pairing correlations, particularly in the weakly bound and continuum regions, which may affect the detailed shell structure and related microscopic properties. A self-consistent inclusion of pairing would require an extension to an RMF+BCS or relativistic Hartree–Bogoliubov treatment in coordinate space. Such an extension is compatible with the present framework and will be considered in future work.}
		
		\section{Acknowledgments}
		This work was supported in part by the National Natural Science Foundation of China No. 12475149, Guangdong Basic and Applied Basic Research Foundation (Grant  No: 2024A1515010911), and the China Nuclear Industry United Graduate School 2025 Special Research Fund (Project Number: HC2025007).

			
	
\bibliographystyle{apsrev4-2}  
\bibliography{ref} 
	
\end{document}